# Principle of Terahertz Time-Domain Spectroscopy

Jiayu Zhao

*Abstract* — In this work, the detection of terahertz (THz) pulse via different schemes of THz time-domain spectroscopy (THz-TDS), including electro-optic sampling (EOS) and photoconductive sampling (PCS), has been reviewed by insisting on its principles and characteristics. Furthermore, four developments of THz-TDS, i.e. THz air-breakdown/biased coherent detection (THz-ABCD), THz-radiation-enhanced emission of fluorescence (THz-REEF), single-shot THz-TDS, and THz asynchronous optical sampling (THz-ASOPS) have also been introduced.

## 1. Introduction

Terahertz (THz) wave, commonly defined from 0.1 THz to 10 THz, is one of the least-explored spectral regions in the electromagnetic (EM) spectrum (between microwave and far infrared), mainly due to the lack of efficient time-domain detection techniques[1]. However, since THz time-domain spectroscopy (THz-TDS) was intensively studied in the 1990s[2], THz science and technology has become a passionate research field receiving increasing attention from all over the world in the past three decades[3]-[10].

Triggered by its cutting-edge applications, THz-TDS has now been broadly applied into diverse fields, such as bio-medicine, material identification, non-destructive evaluation and nonlinear spectroscopy, etc[3]-[10]. THz-TDS is becoming increasingly important and popular because of its broad-bandwidth detection and ability to simultaneously extract both THz amplitude and phase information, which affords a powerful technique for the research of the rich physical and chemical processes in THz range[11],[12].

A THz-TDS setup is, in principle, a coherent (single-cycle) THz emission and detection system. The time-resolved measurement of THz wave is achieved by splitting laser pulses into pump beam to excite pulsed THz radiation and probe beam to temporally sample the THz pulse (by mixing the THz and probe pulses in a detector). The THz temporal waveform thus can be obtained by varying the relative time delay between the THz wave (pump beam) and the probe beam. The detected THz signal is in the form of electric field and its Fourier transformation gives rise to both amplitude and phase information over a wide THz spectral range.

In this work, we have presented the principles of THz-TDS, as well as an overview of six representative experiment schemes of THz-TDS. The basic principles and experimental techniques of THz-TDS have been described in Section 2.1, showing two most widely used systems, i.e. electro-optic sampling (EOS[13]-[15], in Section 2.2) and photoconductive sampling (PCS[16], in Section 2.3). The relevant materials and devices, including EO crystals, PC substrate materials, and optical and electronic devices, have also been discussed. In Section 2.4, the characteristics of THz-TDS setup have been addressed, with emphasis on its performances in terms of sensitivity, signal-to-noise ratio (SNR) and dynamic range (DR). In the same section, THz-TDS and the traditional Fourier transform infrared spectroscopy (FTIR)[17] have been compared, as well. In addition to the standard EOS and PCS based THz-TDS, four developments on THz-TDS, namely, THz air-breakdown/biased coherent detection (THz-ABCD)[18], THz-radiation-enhanced emission of fluorescence (THz-REEF)[19],[20], single-shot THz-TDS[21],[22], and THz asynchronous optical sampling (THz-ASOPS)[23]-[26] have been briefly introduced in Section 3.

## 2. THz time-domain spectroscopy (THz-TDS)

The pioneering studies of materials or gases with THz pulse were initialized in the late 1980s[27],[28], and afterwards this advanced experimental technique was started to be described as the well-known THz-TDS. Recently, THz-TDS has been widely used to investigate the THz applications in fields of physics, chemistry, biology and medicine[29]. The main advantage of THz-TDS relies on its simultaneous measurement of both the amplitude and

Manuscript received ??????, 20??; revised ???????, 20??. This work was supported by ??? under Grant No. ???. Please give the supported projects.

J.-Y. Zhao is with Terahertz Technology Innovation Research Institute, Terahertz Spectrum and Imaging Technology Cooperative Innovation Center, Shanghai Key Lab of Modern Optical System, University of Shanghai for Science and Technology, Shanghai 200093, China (Corresponding author e-mail: zhaojiayu@usst.edu.cn).

??? is with ???, ??? University, ???, ??? (e-mail: ???).

phase of the THz signals transmitted, reflected or scattered by the samples, with broadband response and good performance of signal-to-noise ratio (SNR).

## 2.1 Basic principle

A conventional THz-TDS system consists of a femtosecond laser system, from which the same laser pulse is used for both generation and detection of THz wave. As shown in Fig.1, a femtosecond laser pulse is split into two paths. One was the pump beam and the other was the probe.

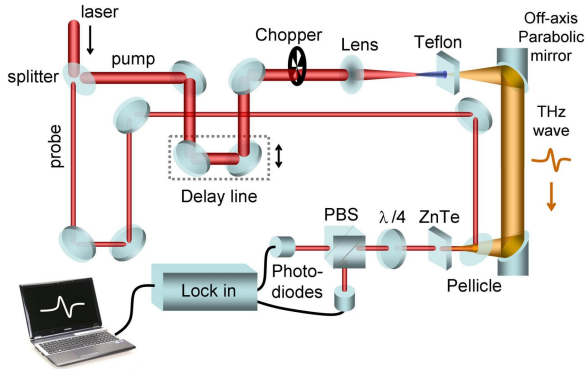

Fig. 1. Schematic diagram of a representative THz-TDS system. A Teflon plate, which has high transmission for THz pulse, is put after the plasma filament to block the residual pumping laser. PBS is short for polarization beam splitter. The other optical and electronic devices are described in detail in the main text.

The pump pulse is used to generate THz wave by means of laser plasma filament (can also be optical rectification in nonlinear crystal or photoconductive antenna, etc), which is created at the focus by focusing the pump beam with a converging lens. The exiting THz pulse from the filament was first collimated by an off-axis parabolic mirror, and then focused by another identical parabolic mirror onto the 'THz detector' (described in detail in Section 2.2 and 2.3).

The probe beam was combined with THz pulse by a Pellicle beam splitter, performing THz-TDS measurement. The basic principle of THz-TDS is to slowly sample a fast THz transient with a probe pulse by coherent detection in time domain, as depicted in Fig. 2.

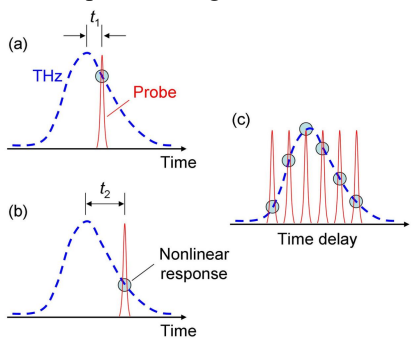

Fig. 2. Schematic of the sampling of a THz temporal waveform with probe pulses via nonlinear response and variable time delay.

The temporal duration of the probe pulse (also the integration time of the detector) is typically much shorter than that of the THz pulse, and hence can be employed as a small temporal 'gate' to stepwise sample the THz waveform. The time-domain sampling is performed when the THz and probe pulses arrive at the detector with a variable delay $t_i$, and generate a corresponding nonlinear response to be detected. This time-variable sampling can be achieved by scanning $t_i$, i.e. by guiding either the pump or the probe beam traveling along a path with an adjustable length, which is generally caused by a spatial movement of a mechanical delay line (with an appropriate step length and range), changing the relative arrival time between the probe and the THz pulses at the detector.

When repeating the measurement of the THz field at the detector for a set of different delays, the full THz waveform $E_{THz}(t)$ could be sampled in its entirety. Fig. 3 provides an example of the measured THz waveform. It could be seen that the THz pulse has the characteristic of single cycle. The corresponding amplitude/power (can also be phase) of each spectral component of the complex THz spectral field can be generated by a following numerical Fourier transformation of the recorded THz temporal data, as shown in the inset of Fig. 3.

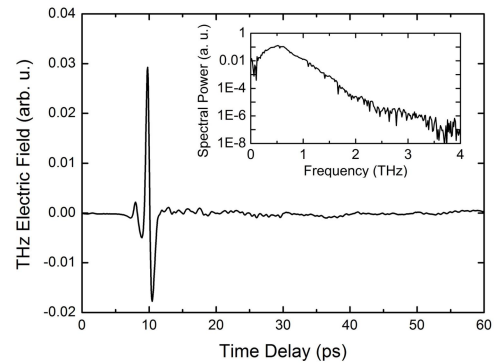

Fig. 3. A representative THz pulse detected by THz-TDS and its Fourier transformation spectrum (inset).

## 2.2 Electro-optic sampling (EOS)

EOS technique is commonly used for the phase-sensitive detection of EM radiation, due to its easy experimental configuration and detectable broad bandwidth. In 1995 to 1996, THz detection by free space EOS was realized due to significant efforts of Nahata et al.[13], Jepsen et al.[14] and Wu et al.[15]. In a THz-TDS system, EOS setup is a widely used 'THz detector', which can be used to detect fast THz transients in a nonlinear EO crystal (such as ZnTe) based on its birefringence (EO effect) induced by the incident THz radiation.

### 2.2.1 Principle

EOS[30],[31], the most common detection method used in the THz community, is realized through an optical rectification process inside a nonlinear EO crystal, such as ZnTe or GaP. The use of EO crystal for EOS typically employs the EO (or Pockel's) effect. When the THz beam interacts with the probe beam inside the EO crystal, the THz electric field induces an instantaneous birefringence in the crystal (Fig. 4 and 5).

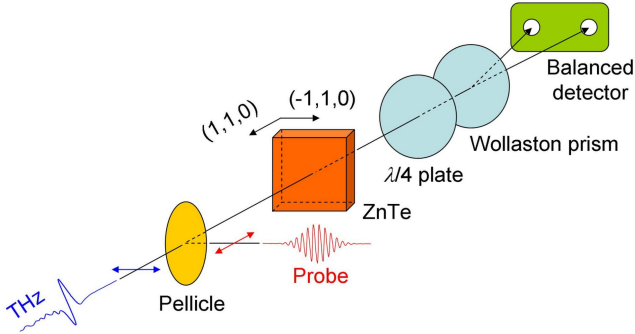

Fig. 4. Schematic diagram of a representative EOS setup.

This birefringence (difference in the refractive indices along each EO crystal axis) leads to a change in the polarization state (polarization rotation, $\propto E_{THz}$) of the collinearly propagating probe pulse, which can be detected after the crystal using a combination of polarization optics ($\lambda/4$ plate + Wollaston prism/polarization beam splitter) and a pair of photo detectors (or a single balanced detector). The use of such 'balanced' detection scheme improves the sensitivity by cancelling most of the laser fluctuations and achieves better SNR. As for the time-resolved detection of the THz pulse $E_{THz}(t)$, the transient birefringence is sampled with the probe pulse at different time delays.

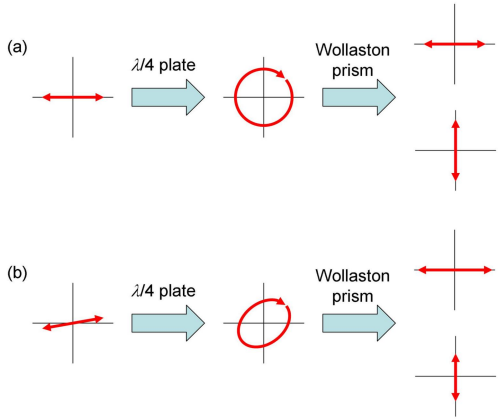

Fig. 5. The horizontal polarization of the probe pulse (a) is rotated via EO effect (b) when THz pulse co-propagates inside the EO crystal, which will result in a final non-zero output of the balanced detector.

### 2.2.2 EO crystal

The performance of EOS is limited by the dielectric properties of the EO crystal. Since LiTaO₃ crystal was first demonstrated as the EOS element in 1995 by Wu and Zhang[15], a variety of EO materials without inversion symmetry, including semiconductors, inorganic and organic crystals, have been tested as 'detector' for applications in THz-TDS[32],[33]. Among such crystals, ZnTe and GaP are most commonly used for EOS because of their large EO coefficients and good transparency at the wavelength of both the incident THz radiation and probe beam. Apart from ZnTe and GaP, LiTaO₃, GaAs and DAST are also excellent candidates for EOS of the THz pulse. Properties of these EO crystals have been compared in Table 1, including EO coefficient, group refractive index and surface orientation.

Table 1. Properties of typical EO crystals[32],[33]

| EO crystal | EO coefficient (pm/V) at ($\mu$m) | Group refractive index at ($\mu$m) | Surface orientation |
|---|---|---|---|
| ZnTe | $r_{41}$ = 4.04 at 0.633 | 3.224 at 0.835 | 110 |
| GaP | $r_{41}$ = 0.97 at 0.633 | 3.556 at 0.835 | 110 |
| LiTaO₃ | $r_{33}$ = 30.5 at 0.633 | — | — |
| | $r_{13}$ = 8.40 at 0.633 | | |
| GaAs | $r_{41}$ = 1.43 at 1.15 | 3.61 at 0.886 | — |
| DAST | $r_{11}$ = 160 at 0.820 | — | — |

An important condition for broadband detection of THz wave by EOS is the EO crystal's optical phonon absorption band, which is insensitive to the incident THz wave due to the phonon resonances. Compared with ZnTe, GaP has a higher-frequency optical phonon absorption band[32] and thus can be used to achieve broader-bandwidth detection of THz radiation. However, the EO coefficient of GaP crystal is only one fourth that of ZnTe[32].

Another crucial factor is the appropriate phase matching between the THz and probe pulses, which propagate on the same axis inside the EO crystal. Phase matching is achieved when the phase velocity of THz wave is equal to the group velocity of probe pulse (or the velocity of probe envelope)[34]. For an EO crystal with dispersion at optical frequencies, the phase difference $\Delta\Gamma$ of the probe beam is proportional to the crystal thickness $d$ as described by[34]:

$$\Delta\Gamma = \frac{2\pi}{\lambda} d n_g^3 r_{41} E_{THz},\qquad(1)$$

where $n_g$ is the group refractive index at the wavelength $\lambda$ of the probe beam, and $r_{41}$ is the EO coefficient of the EO crystal.

Due to the difference in the respective dispersions, the phase mismatch accumulates after the propagation of the two waves through the crystal. Therefore, a thinner EO crystal normally detects higher THz frequencies than a thicker one, because of the phase mismatch (shorter coherence length) and also stronger optical phonon absorption in the latter. If possible, the EO crystal thickness should be estimated before applying EOS, by taking into account of the phase-matching condition dependent on the target THz frequency range, as well as the probe central wavelength.

### 2.2.3 Optical and electronic devices

To develop a THz-TDS, one requires both optical and electronic elements. This section discusses some aspects of these devices.

*Off-axis parabolic mirror*

Off-axis parabolic mirrors are used to collimate and focus the THz radiation from source to detector, or focus the THz wave to a diffraction-limited spot on a potentially small sample. Its advantages are (i) large area, (ii) high reflectivity (> 95 % in entire THz range with protected gold coating), and (iii) no echo pulses.

*Chopper and lock-in amplifier*

During EOS, the detected THz signal is integrated over a large number of pulses with a lock-in amplifier. Accordingly, a mechanical (optical) chopper is applied to periodically modulate the pumping laser pulses (or the bias on the photoconductive emitter). The function of a lock-in amplifier is as follows: when the THz signal is chopped, the lock-in amplifier detects the fraction of the modulated THz signal with the same frequency, canceling out most other frequencies and DC parts. Therefore, weak THz signals on a high background noise level can be detected.

*Optical delay line*

The optical delay line enables the ability to scan the time delay. The used optical delay line is normally a combination of two mirrors orthogonally mounted on a stepper motor driven device with high mechanical stability and good positioning accuracy. So it is possible to measure small change of time difference $\Delta t$.

On the other hand, the maximum scanning range (time window $T_{window}$) is of importance for spectral resolution $\Delta v$ (the inverse of $T_{window}$) of the THz-TDS system, given by the following mathematical relation:

$$\Delta v = \frac{1}{T_{window}} = \frac{1}{N \cdot \Delta t},\qquad(2)$$

where $N$ is the number of data points recorded in the time domain. As an example, typically, THz-TDS systems employ a 3-cm-long delay line, leading to $T_{window}$ = 200 ps and $\Delta v$ = 5 GHz. This is the reason for scanning long time delay and recording the THz signal over a large $T_{window}$ during EOS in order to increase the spectral resolution. However, longer delay line requires a perfect alignment of the laser beams, as well as an accurate control of the laser power stability, which need to be taken into account.

Until very recently, the EOS technique has been the only method for the ultra-broadband detection of the THz radiation. Compared with EOS, although the photoconductive antenna (PCA) was a key device to initiate the recent development of THz science and technology, the photoconductive sampling (PCS) method has become somewhat old-fashioned. However, PCS is still widely adopted in THz labs due to its easy and integrated experimental configuration.

### 2.3 Photoconductive sampling (PCS)

The first report of THz-TDS for the detection of THz radiation was by Auston *et al.*[16] in 1984 via PCA, which is the earliest THz pulse detector, typically containing a pair of micro-fabricated metal electrodes with a small gap on a semiconductor substrate[35]. A sensitive electronic current amplifier is used to connect both electrodes[36]. When a PCA based PCS setup is used to detect the THz radiation, the basic detection principle and some related characteristics are described as follows.

### 2.3.1 Principle

During PCS, a PCA acts as the THz detector, as shown in Fig. 6, which is triggered by a time-delayed probe laser pulse to generate photo-created carriers in the electrode gap. This time delay is adjusted thanks to an optical delay line, as described in Section 2.2.3.

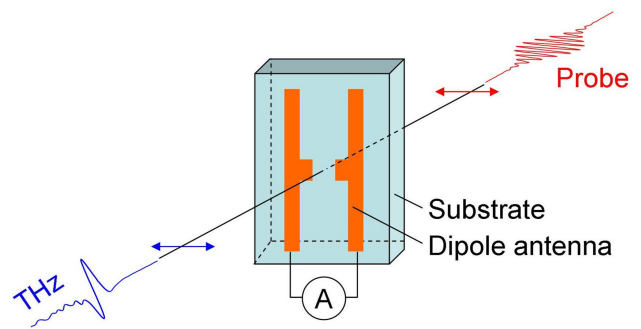

Fig. 6. Schematic diagram of a representative PCS setup.

The incident instantaneous THz field coupled by the metallic dipole antenna provides a local bias field between two electrodes and accelerates the mobile carriers, creating transient photocurrent, which is then measured by a current meter (Fig. 7). In order to optimize the THz detection of PCS scheme, the polarization directions of both the THz and probe pulses need to be perpendicular to the edge of electrodes[37], as displayed in Fig. 6. In this way, the measured THz pulse will retain the characteristic of single cycle[37] with better performance of detectable bandwidth and signal-to-noise ratio (SNR).

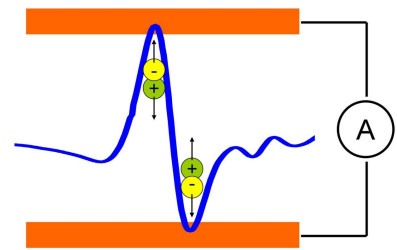

Fig. 7. Between two electrodes, the THz electric field accelerates the carriers and creates transient photocurrent, which can be detected by a sensitive current meter.

The measured photocurrent $J(t)$ induced by the incident THz radiation at a time delay $t$ is proportional to the product of the incident THz electric field $E_{THz}(t)$ and the number density of the existing photo-created carriers $N(t)$[38]:

$$J(t) = e\mu \int_{-\infty}^{\infty} E_{THz}(t') \cdot N(t'-t)dt',\qquad(3)$$

where $e$ and $\mu$ are elementary electric charge and electron mobility, respectively. Noting that, for simplicity, $N(t)$ actually includes the convolution of the probe pulse intensity and the carrier density.

By controlling the time delay between THz and probe pulses, the temporal change of $J(t)$, which is proportional to the incident THz electric field, can be sampled with a current meter. Compared with the balanced detection manner of EOS, the THz electric field to be PC sampled is directly converted to the photocurrent in a PCA, which means that PCA is as robust to the optical noises as EOS technique.

From Eq. (3) one can also see that, when the PCA is considered as a sampling detector, the temporal increase and decrease of $N(t)$ should be as short as possible (best if $N(t)$ were a $\delta$-function), so that $J(t)$ would exactly reflect $E_{THz}(t)$. In reality, $N(t)$ will be restricted by several factors, such as probe pulse width and carrier lifetime of the PCA substrate material.

Clearly, the probe pulse duration must be much shorter than the photo-excited carrier lifetime, and the latter must be also significantly shorter compared to the THz pulse period, in order that the PCA response to the THz field is almost instantaneous, and the measured THz field has sufficient time resolution. Otherwise, the PCA will qualitatively work as an integrating detector according to Eq. (3). This is one of the reasons to explore the substrate materials with short carrier lifetime.

### 2.3.2 Substrate material

Performance of a PCA depends mainly on the substrate material, whose carrier lifetime essentially limits the bandwidth of the detected THz emission. Especially, materials with a short carrier lifetime are usually selected as substrate in order to increase the response speed of PCA and achieve broadband detection.

With the motivation to shorten the carrier lifetime, low-temperature-grown GaAs (LT-GaAs) and silicon-on-sapphire (RD-SOS) has been well developed[39]-[42]. The typical lifetime of LT-GaAs is around 0.6 ps[39],[40]. An ultra-broadband THz detection with frequency components over a few tens of THz has been demonstrated in previous reports[41],[42] with a LT-GaAs based PCA.

### 2.3.3 Frequency response

Apart from the carrier lifetime of the substrate material,

probe pulse width is another crucial factor influencing the frequency response of a PCA. In case of PCA working as a PCS detector, it can sufficiently operate in low THz frequency regimes, generally below 5 THz. The frequency response of PCA in the low THz band is almost independent on the probe pulse width. On the other hand, the decay of the high-frequency components of the THz spectra critically depends on the pulse width of the probe beam. As the pulse width of the probe beam was broadened, the high THz band rapidly decreases (even faster than the EOS spectrum).

The significant influence of probe pulse width on PCA frequency response was confirmed by Ref. [41], which has demonstrated that the detectable bandwidth of THz radiation could be extended up to 60 THz with PCA sampled by the probe pulse width of only 15 fs. In order to obtain broadband detection of THz wave, the pulse duration of the probe beam should be as short as possible for both PCS and EOS. Compared with EOS, the frequency response of PCA, mainly dependent on the probe pulse width, is advantageous in the spectral analysis.

### 2.4 Characteristics

In this section, we will investigate some characteristics that affect the overall performances of the THz-TDS system.

### 2.4.1 Sensitivity

One of the most attractive features of THz-TDS is the ability to coherently detect both the THz field amplitude and phase[43], which results in a far higher sensitivity compared to intensity detection (e.g. with a thermal detector or conventional photodiode). This is because the measured signal $S_{THz}$ is proportional to the THz field amplitude, i.e. $S_{THz} \propto E_{THz} \propto \sqrt{I_{THz}}$ as opposed to the intensity detection signal $S_{THz}' \propto I_{THz}$ in case of an intensity detector. Therefore, if the THz intensity/power is reduced by a factor of 100, the coherently detected signal only decreases by a factor of 10.

### 2.4.2 Signal-to-noise ratio (SNR)

The SNR of the THz waveforms measured by the EOS and PCS techniques are similar to each other. The estimated SNR in the EOS method with the same electric field of the THz radiation is about four times larger than that in case of PCS is applied[44].

Recalling that SNR is defined by:

$$SNR(\omega) = \frac{S(\omega)}{\sigma(\omega)},\qquad(4)$$

where $S(\omega)$ is the THz amplitude at angular frequency $\omega$, $\sigma(\omega)$ is its standard deviation. The design of PCA might have a complicated impact on $\sigma(\omega)$, including the collecting efficiency of THz field by metallic antenna, the electrical properties of the semiconductor substrate (dark resistivity, carrier mobility and lifetime) and the excitation conditions (photo-carriers density), which leads to worse SNR of PCS than EOS. Normally, one has to select a PCA with fast response to increase the SNR of a PCS setup.

### 2.4.3 Dynamic range (DR)

If $\sigma(\omega)$ in Eq. (4) is THz signal independent, then $\sigma(\omega)$ will equal to $S_{min}(\omega)$, which is the minimum measurable THz signal at $\omega$. Then the formula of SNR will be deduced as

$$DR = \frac{S(\omega)}{S_{min}(\omega)}, \qquad (5)$$

which is the definition of DR.

Since THz-TDS detects THz electric field rather than THz intensity, the measurement outcomes typically have greater DR than conventional techniques. As an example, considering an optically dense medium with a very small power transmission coefficient of $10^{-8}$, a sensitive THz-TDS system can still characterize this medium by measuring the THz field transmission coefficient of $10^{-4}$.

### 2.4.4 Comparison with Fourier transform infrared spectroscopy (FTIR)

Far infrared spectral properties of materials have long been studied for more than 100 years[7]. Until the 1980s, the most powerful tool for spectroscopic analysis in the far-IR regime with wavelength between 20 and 500 μm was FTIR, which is based on a black body source (thermal emitter), a Michelson interferometer and a bolometer[17]. By contrast, THz-TDS was just proposed about 30 years ago. However, THz-TDS has become more and more useful in the past three decades.

Compared with FTIR, it has been recognized that the most important advantage of THz-TDS is the possibility of measuring not only the amplitude of the THz signal, but also its phase. This allows determining the complex refractive index of a sample, which can avoid the uncertainty caused by the Kramers–Kronig analysis.

On the other hand, based on THz-TDS, the THz signal is observed in the form of a time trace with sub-picosecond time resolution, and the coherent nature of THz detector dramatically reduces its minimum detectable power. Thus THz-TDS allows one to perform time-of-flight and time-resolved measurements with better sensitivity than FTIR.

In terms of bandwidth, FTIR is of course better than THz-TDS as it can be used from the visible to the far infrared domain. In general, THz-TDS has a better SNR than FTIR below 3 THz, whereas FTIR is preferable over 5 THz[45]. Therefore, before widely replaced by THz-TDS, the traditional FTIR remains a mature and popular approach, which employs reliable, well established and easy-to-use apparatuses.

## 3. Progress on THz-TDS

In this section, four developments of THz-TDS have been briefly introduced, including THz air-breakdown/biased coherent detection (THz-ABCD), THz-radiation-enhanced emission of fluorescence (THz-REEF), single-shot THz-TDS, and THz asynchronous optical sampling (THz-ASOPS).

### 3.1 THz-ABCD

Detection of THz wave with laser-induced air plasma as not only THz wave emitter but also THz detector was initially reported in 2006[18]. The basic mechanism for using gaseous media to sense THz waves primarily lies in the THz-induced second harmonic generation through a third-order nonlinear process[18],[46],[47], as shown in Fig. 8.

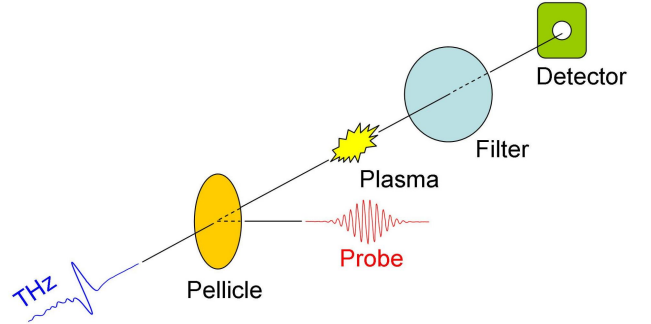

Fig. 8. Schematic diagram of a THz-ABCD setup. A filter is put after the plasma filament to block the residual probe laser. The detector is used to measure the time-resolved second harmonic signal induced by mixing the probe and the THz pulses.

The interaction between $\omega$ pulse (probe) and THz wave in laser-induced air plasma results in $2\omega$ pulse, which can be described in the following equation:

$$E_{2\omega} \propto \chi^{(3)} E_{\omega} E_{\omega} E_{THz}, \qquad (6)$$

where $E_{2\omega}$, $E_{\omega}$, and $E_{THz}$ are the electric field amplitudes of the $2\omega$, $\omega$, and THz waves, respectively, and $\chi^{(3)}$ is the third-order susceptibility of air. Since $E_{2\omega} \propto E_{THz}$, the intensity of the measured second harmonic is proportional to the intensity of the THz wave, i.e., $I_{2\omega} \propto I_{THz}$. This technique has been named as THz air-breakdown coherent detection (THz-ABCD).

In order to further reduce the probe pulse energy required for THz detection of this all-air based THz-TDS method, an external AC bias across the probe-induced plasma volume has been applied. Correspondingly, THz-ABCD changes its meaning from "THz air-breakdown coherent detection" to "THz air-biased coherent detection".

Since air is the only medium used for THz detection, THz-ABCD has unique features, such as no phonon absorption or significant phase-matching issue over THz band as those existing in EO crystal. Therefore, only the probe pulse duration could limit the detection bandwidth of THz-ABCD, which is obviously much broader than that of EOS. The useful spectral range of a typical THz-ABCD system continuously covers from 0.2 to over 30 THz[48].

## 3.2 THz-REEF

The interaction between THz wave and femtosecond laser-induced air plasma is of great importance to the THz pulse detection, since it not only can result in second harmonic generation (as described in the previous section of THz-ABCD), but also can significantly enhance the fluorescence emission from the plasma, which has been demonstrated as an omnidirectional THz-TDS method[19],[20], named as THz-REEF. In THz-REEF experiment as shown in Fig. 9, a single-cycle THz pulse is collinearly focused to the focal region (air plasma) of the laser beam, and the fluorescence emission is detected by a combination of monochromator and photomultiplier tube (PMT).

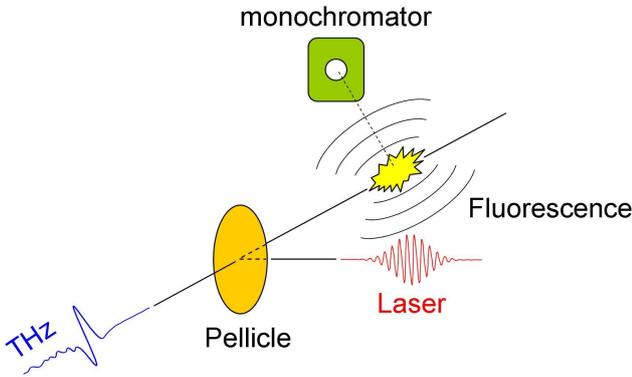

Fig. 9. Schematic diagram of a representative THz-REEF setup.

Inside the plasma, the irradiation of THz pulse leads to an increase of plasma temperature through electron acceleration in the THz field, and thus the total kinetic energy of the electrons is increased. On the other hand, inside the plasma, there exists many high-lying states of atoms and molecules formed via absorption of multiple laser photons[49]. Compared to the ground states, they are more easily ionized by inelastic collisions with the surrounding energetic electrons. Therefore, the energy transferred from THz wave to electrons would increase the rate of electron-impact ionization of the highly excited particles, and larger ion population will lead to more fluorescence emission. As a result, the fluorescence emission is enhanced by the THz field.

In this manner, the THz waveform information is encoded into the change of the fluorescence at different time delay between THz and laser pulses. The enhanced fluorescence $\Delta I_{FL}$ can be approximate as follows:

$$\Delta I_{FL} \propto \int E_{THz}^2(t)dt \,, \tag{7}$$

which shows the agreement between the THz intensity and derivative of the enhanced fluorescence. Thus, the time-dependent fluorescence contains the information of the time-resolved THz intensity. Due to the high transparency and omnidirectional emission pattern of the fluorescence (ultraviolet, UV), THz-REEF can be used to measure THz pulses at remote distances, e.g. 10 m[20], with minimal water vapor absorption and unlimited directionality for THz signal collection.

## 3.3 Single-shot THz-TDS

The conventional THz-TDS uses a mechanical delay line to vary the optical path between the pump and probe pulses[50],[51], and the THz electric field is repetitively sampled for each sequential time delay. Therefore, the data acquisition rate via this temporal scanning is very limited, which clearly cannot meet the requirement for real-time THz-TDS measurements.

To increase the acquisition rate, one possible method is based on single-shot THz-TDS[21],[22], whose most important feature is the capability of the measurement of a full THz temporal waveform with only one chirped probe pulse (without using mechanical delay line). This will significantly speed up the acquisition rate.

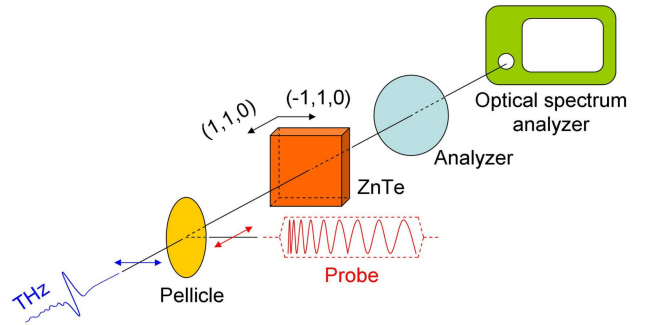

Fig. 10. Schematic diagram of a representative single-shot THz-TDS setup. The polarization of different wavelength components of the chirped probe pulse is rotated by different portions of THz electric field through the Pockels effect inside the EO crystal. More precisely, the degree and direction of the probe polarization rotation is proportional to the THz field strength and polarity. After the optical analyzer, this polarization modulation is converted to the amplitude variation on the optical spectrum analyzer (OSA).

As schematically illustrated in Fig. 10, a representative single-shot THz-TDS system is similar to the aforementioned EOS setup, except for the use of a chirped probe pulse and an optical spectrum analyzer (OSA). As for the probe pulse, it has been linearly frequency chirped (a linear ramp of frequency versus time) and time stretched to tens of picoseconds. Then, the chirped probe pulse and the THz pulse co-propagate in the EO crystal.

Inside the EO crystal, one can see from Fig. 11 (Right) that, the information of THz temporal waveform is linearly encoded onto the different spectral components of the chirped probe pulse, and finally decoded by dispersing the probe beam in an OSA (Fig. 10).

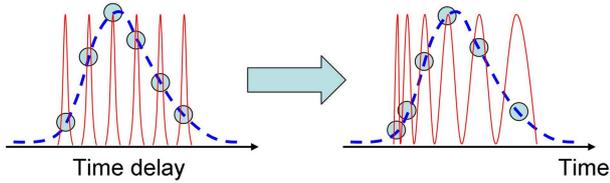

Fig. 11. Schematic of traditional (Left) and single-shot (Right) THz-EOS. Different wavelength component of the chirped pulse is modulated by different portion of the THz electric field via EO effect.

The THz temporal waveform can be extracted by performing a subtraction between the probe spectra with and without THz pulse modulation (as background). The resulting differential spectral distribution reconstructs both amplitude and phase of the THz pulse.

### 3.4 THz- ASOPS

A significant disadvantage of the conventional THz-TDS system is the low acquisition rate for THz transients with high spectral resolution. For example, the detection of a 1-ns-long THz temporal waveform (corresponding spectral resolution of 1 GHz) requires a travelling distance of 2×15 cm of a mechanical delay line, which is usually limited in scanning speed. Therefore, the total data acquisition time could be in the range of a few (or even tens of) minutes.

This tradeoff between the THz spectral resolution and data acquisition rate can be avoided by asynchronous optical sampling (ASOPS)[23], which is a THz-TDS technique introduced in the late 1980s without using a mechanical translation stage.

As shown in Fig. 12, THz-ASOPS[24],[25] is based on two separate femtosecond laser sources with constant repetition frequency of $(f + \Delta f)$ and $f$, respectively, close to each other due to the slight offset frequency $\Delta f$. One laser serves as pump to generate the THz radiation, and the other is probe used for EOS of the THz waveform.

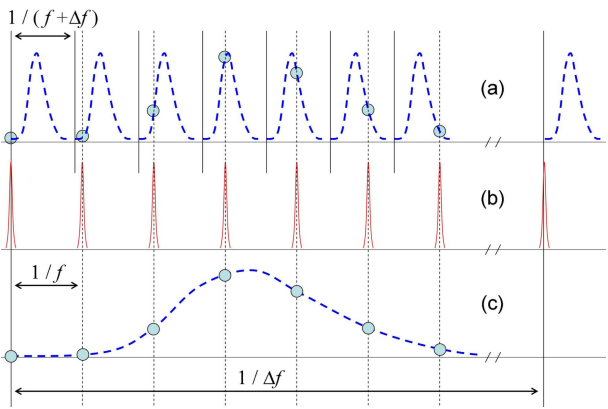

Fig. 12. Sketch of the THz-ASOPS principle. (a) The pump laser generates THz pulses at a repetition frequency of $f+\Delta f$, which are sampled by (b) the probe with $f$. (c) The sampled THz waveform.

In such a scheme, the time delay between pulse pairs from the two lasers is repetitively scanned (linearly ramped) between zero and the inverse laser repetition rate (i.e. $1/f$) at a scanning rate given by $\Delta f$. That is, the available time window $T_{window}$ is $1/f$ (thus the spectral resolution is $f$, according to Eq. (2)), whereas the data acquisition rate equals to $\Delta f$.

Generally, laser repetition rate of $f = 1$ GHz and $\Delta f$ in the kHz frequency range[26] (e.g. 10 kHz) are favorable for ASOPS, leading to the detected THz spectral resolution as high as 1 GHz with data acquisition time of only 0.1 ms. This performance of ASOPS is impossible with traditional THz-TDS systems with mechanical delay stages.

## 4. Conclusion

In summary, in the present article, we have reviewed the basic principles and experimental techniques of THz-TDS. Six representative schemes (EOS, PCS, THz-ABCD, THz-REEF, single-shot THz-TDS and THz-ASOPS) have been discussed, in terms of principles, experimental performances, characteristics, and relevant materials and devices. In addition, a brief comparison between THz-TDS and FTIR has also been made.

### Acknowledgment

Authors acknowledge the support of ???.